\documentstyle[12pt]{article}

\begin{document}

\pagestyle{empty}
\renewcommand{\thefootnote}{\fnsymbol{footnote}}
\def\lsim{\:\raisebox{-0.5ex}{$\stackrel{\textstyle<}{\sim}$}\:}
\def\gsim{\:\raisebox{-0.5ex}{$\stackrel{\textstyle>}{\sim}$}\:}

\def\bib{\bibitem}
\def\be{\begin{equation}}
\def\ee{\end{equation}}
\def\beqar{\begin{eqnarray}}
\def\eeqar{\end{eqnarray}}
\def\barr{\begin{array}}
\def\earr{\end{array}}
\parindent=20pt

\bigskip

\begin{center}

{\large \bf DECAY-TIME ASYMMETRIES AT THE B-FACTORIES\footnote{Invited
talk at the X DAE High Energy Physics symposium in December 1992, held at
TIFR, Bombay.}}

\bigskip
\bigskip

K. V. L. SARMA\\
Tata Institute of Fundamental Research\\
Homi Bhabha Road\\
Bombay 400 005.\\

\bigskip
\bigskip

1.~INTRODUCTION
\end{center}

\medskip

The primary motivation for proposing to build a $B$-Factory is to explore
the signals for CP violation in the $B$ meson decays. The idea is to
establish a rate asymmetry which is implied by unequal rates for $B$ to
decay into a state $f$ and $\bar B$ to decay into the CP-conjugate state
$\bar f$,
\be
\Gamma (B \rightarrow f)  \not= \Gamma (\bar B
\rightarrow  \bar f).
\ee
One might look for such a violation in the neutral mesons $B_d$ and $B_s$, and
in the charged ones $B_u$. Here I shall focus mainly on the neutral
system $B_d - \bar B_d$, and sometimes refer to it as $B-\bar B$. But
before I discuss some of the asymmetries and the importance of their
measurement, let me say a few words about the ``Asymmetric $B$-Factories'',
mainly for pedagogical reasons.

{\bf Why Factory?}: ~Almost all the hadronic decay modes of the $B$ meson
could be called ``rare'' modes because observations show that each of them
has a decay rate which is quite small compared to the total rate. For
instance, according to the 1992 data of the CLEO group [1] we have the
branching fractions (BF)
\begin{eqnarray}
BF(B_d \rightarrow J/\psi  + K_S) & = & (1.02 \pm 0.43 \pm 0.17)
\times 10^{-3}, \\
BF(B_d \rightarrow \pi^+ \pi^-)  & \lsim & 4.8 \times 10^{-5}, (90\% CL);
\end{eqnarray}
the first decay mode is CKM-allowed and the second one is CKM-suppressed.
This kind of numbers is perhaps not unexpected: in the decay of a meson of
5 GeV mass, kinematics would admit many distinct hadronic channels and,
treating all of them on an equal footing, each gets only a
tiny fraction of the total width. To increase the statistics therefore one
would have to start with a large sample of B's and hence the need for a
``$B$-Factory''.

The Factory is nothing but a high luminosity $e^+ e^-$ collider which is
tuned to produce the well-known onium-state $\Upsilon (4S)$ at mass 10.580
GeV. This $b\bar b$ resonance is produced with a cross section of
$1.15\;nb$, and lies only about 23 MeV above the $B\bar B$ threshold. It
decays into the two hadronic modes, $B^+B^-$ and $B_d^0 \bar B_d^0$, each
with a $BF$ very close to 50\%;
\be
e^+e^- \rightarrow \Upsilon(4S) \rightarrow B\bar B.
\ee
To get an idea of the peak luminosities which are being talked about we
quote the design value for the Cornell Factory CESR-B [2],
$L(peak)= 3 \times 10^{33} cm^{-2} s^{-1}$; for comparison, recall the
CESR 1992 peak value $2.5 \times 10^{32} cm^{-2} s^{-1}$, which is the
present world record for the storage rings.  This means an
integrated luminosity of about $100\;fb^{-1}$ in about 3 years of running,
at an average $L$ equal to a third of the peak value (the integrated value
of 1991 was $1.2\;fb^{-1}$).  Note that values like $100\;(fb)^{-1}$ are
needed to study CP violation as they yield typically about $10^{8}$ events
of interest.

The second feature in $B$-Physics experiments pertains to the measurements
of small decay times which lie in the picosecond range. The $B$ meson
lifetimes determined from the 1992 LEP (averages of the ALEPH and DELPHI)
data, are [3]
\begin{eqnarray}
\tau (B_d) & = & 1.44 \pm 0.18\;ps, \nonumber \\
\tau (B_u) & = & 1.33 \pm  0.19\;ps, \\
\tau (B_s) & = & 1.05 \pm  0.31\;ps. \nonumber
\end {eqnarray}
As the $B_d$ has a speed $\beta^* = 0.065$ in the $\Upsilon (4S)$ rest
frame, it travels a very short distance before decaying; the average decay
length with the lifetime quoted above (ignoring the error) is
\be
d^* = \gamma^*\beta^* c\tau = 28\,\mu m.
\ee
When the colliding $e^+$ and $e^-$ have unequal energies it corresponds to an
`Asymmetric' $B$-Factory in which the $\Upsilon$ moves.
The $B$'s from such a moving
source can then travel larger distances before they decay. Note that it is
the increased speed of the $B$ that helps here, and not so much the Lorentz
time-dilatation.

To be specific, the Cornell Factory CESR-B, [2] will be a two-ring collider
with 8~GeV $e^-$ hitting 3.5~GeV $e^+$. The Upsilon will be produced with a
velocity $\beta = 0.39$ along the direction of the $e^-$.
The resulting $B$'s would have boost factors in the range
\be
\beta\gamma = 0.36 - 0.50,
\ee
as the emission angles vary, and hence they would have larger decay distances
\be
d = 153 - 215\, \mu m.
\ee
Measurements of such lengths are within the easy reach of the present day
vertex detectors using silicon micro-strips.

The intervals of time, or corresponding lengths, which one might be
interested in resolving at the Factory could be smaller still:~The actual
structure of the colliding bunch typically will have a small transverse
cross section of ($50\;\mu m \times 50\;\mu m$) but a large longitudinal
dimension of about $1.5\,cm$.  This kind of a needle-shaped bunch makes it
hard to locate the point of creation of the $\Upsilon$ resonance with the
needed accuracy at the asymmetric Factory; nonetheless one could ascertain
easily the `gap' between the two decay-instants $t$ and $t'$ of the
boosted $B\bar B$pair. Hence one expects to have a good knowledge of the
relative time
\be
\tau = t - t'.
\ee
As this time-difference could take values much smaller than a picosecond,
one would have to determine the corresponding gaps between decay vertices
which are smaller than $150~\mu m$.

For measuring certain asymmetries we do not need the value of $\tau$; it
is enough if we can ensure that it is positive or negative. Note that at
an asymmetric collider, event reconstruction will be easier than at the
symmetric one because the separation between decay vertices of the two $B$'s
is increased by the boost.  On the other hand too much boost can be
couter-productive because it focuses or squeezes all the produced
particles into a narrow cone and vertexing becomes a technical challenge.

{\bf Bottom Tagging}:  How do we ascertain whether it was a $B$ that decayed
and not a $\bar B$? Recall that in semileptonic decays the $W$ couplings
that govern the decay are: $b\rightarrow W^- c(u)$ and $\bar b \rightarrow
W^+ \bar c (\bar u)$.
Since the charge of the emitted lepton has to be the same as that of the
emitted $W$, the $B$ flavour at the instant of decay can be tagged by the
charge of the primary lepton (say, a muon with at least 1.5 GeV in the
$\Upsilon (4S)$ rest frame):
\begin{eqnarray}
B(\bar b d)& \rightarrow & \ell^+\;\; + \ldots, \nonumber \\
\bar B(b\bar d) & \rightarrow & \ell^-\;\; + \ldots\; .
\end{eqnarray}
The backgrounds for such events arise mainly from the charm and tau
decays, e.g., through the chains $B \rightarrow D \rightarrow \ell$ and
$B \rightarrow \tau \rightarrow \ell$, and they are
estimated by Monte Carlo simulations.

As for hadron tags the inclusive mode with a charged kaon (e.g., due to
the cascade decays $B \rightarrow \bar D + \ldots \rightarrow K^+ +\ldots$)
may be promising because of its
large $BF = (85 \pm 11)\%$.  Monte Carlo studies indicate that background $K$'s
from lighter hadrons might be manageable; e.g. below $3\;{\rm GeV}/c$,
if a $K$ is
detected with 100\% efficiency the $B$ flavour could be determined with about
40\% efficiency which is 3 - 4 times larger than the lepton-tagging
efficiency.

An important feature of the Upsilon factory is the possibility of
measuring an observable which is odd with respect to time; such
measurements are obviously not possible with the ``conventional'' beams of
$B$'s. In the $\Upsilon (4S)$ rest frame let us suppose we have identified the
left-moving bottom meson to be a $\bar B$  at time $t_L$ because the lepton
tag was $\ell^-$. Then its companion moving towards the right must be a $B$ at
time $t_L$.  However the time $t_R$, the instant of decay of the right mover,
could be either later or earlier than $t_L$. In this case therefore we say
that the right-moving $B$ can `evolve' forward or backward in time before
its decay, because the zero of time ($t=0$) at the Factory is decided by the
tag-time.

Summarizing, the Asymmetric $B$-Factory is designed to produce copious
numbers of correlated $B\bar B$ pairs, each pair travelling at considerable
(constant) speed. Information would be available on the relative time
difference between the two decays of a given $B\bar B$ pair, wherein one of the
decays serves as the bottom-tag.

My plan is to talk briefly about the following specific topics:

\begin{description}
  \item [(a)] CP violating asymmetries to test the SM; Winstein ambiguity
  \item [(b)] Data on dilepton charge asymmetry and CP violation
  \item [(c)] Asymmetry for determining the parameter y
  \item [(d)] Validity of the $\Delta B = \Delta Q$  rule.
\end{description}

First a mention of the notation and some general conclusions based on the
box-diagram of the standard model:

{\bf Notation}: Let the propagating states $B_k (k=1,2)$ which have complex
masses $\left( m_k - {i\over 2} \Gamma_k \right)$  be defined as
\begin{eqnarray}
B_1 & = & p B + q \bar B  \nonumber \\
B_2 & = & p B - q \bar B.
\end{eqnarray}
The phase convention is $(CP)\;B = +\bar B$. We define the following parameters
\begin{eqnarray}
g & = & q/p, \nonumber \\
x & = & (m_2 - m_1)/\Gamma, \nonumber \\
y & = & (\Gamma_2 - \Gamma_1)/(\Gamma_2 +\Gamma_1), \nonumber \\
\Gamma & = & (\Gamma_2 + \Gamma_1)/2.
\end{eqnarray}

Let us consider the $B$ decays into CP eigenstates f where
\be
\bar f \equiv (CP) f = \pm f,
\ee
and denote the decay amplitudes by
\begin{eqnarray}
A_f & = & {\rm Ampl} (B \rightarrow f), \\
\bar A_f &  = & {\rm Ampl} (\bar B \rightarrow f).
\end{eqnarray}
We define the
(phase-convention independent) complex parameter
\be
u_f  = g\; {\bar A_f \over A_f},
\ee
which, if CP invariance holds, would take the value $+1~(-1)$ for CP-even
(CP-odd) eigenstates $f$.

{\bf Lore of the ``BOX''}: A study of the standard box-diagram with the
exchange
of $W$-pair, and assuming the number of quark generations to be 3, leads us
to expect the ratio
\be
{|\Gamma_{12}| \over |M_{12}|}\; \simeq {O (m^2_c) \over O (m^2_t)}\; \lsim
10^{-3}
\ee
to be quite small (see e.g., Ref 4). Here, as per the usual notation, the
off-diagonal mass-matrix element $[M_{12} - i \Gamma_{12}/2]$  is
essentially $p^2$; we recall that $\Gamma_{12}$ gets contributions only
from the physically allowed $b$ quark decays. Thus we expect the mixing
ratio $g$ mainly to be a phase factor,
\be
g = {q \over p}= \left[{M^*_{12} - {i\over 2}\;\Gamma_{12}^* \over M_{12}
- {i\over 2} \Gamma_{12}}\right]^{1/2} \simeq \left[{M^*_{12}  \over
M_{12}}\right]^{1/2},
\ee
and we can set
\be
|g| \simeq 1,
\ee
and because $y$ is determined by $Re(\Gamma_{12})$,
\be
y  \simeq 0.
\ee
Although not necessary but for the sake of displaying formulas which are
simple, we shall also assume that
\be
|u_f| \simeq 1.
\ee

\bigskip

\centerline  {2.~CP VIOLATING ASYMMETRIES}

\medskip

Let us start with pure states $B$ and $\bar B$ at proper time $t=0$.  The
time-evolved states $B(t)$ and $\bar B(t)$ would decay into a common channel
$f$ with the rates given by (see Refs. [4-8] and references therein)
\be
{d\Gamma \over dt}\;\left[ \begin {array} {c} B(t) \rightarrow f \\ \bar
B(t) \rightarrow f \end {array} \right] \sim |A_f|^2 e^{-\Gamma t} \left[ 1
\mp Im(u_f)\;\sin (x\Gamma t)\right].
\ee
One constructs the time dependent asymmetry
\begin{eqnarray}
A(t) & \equiv & {d\Gamma (B \rightarrow f) - d\Gamma (\bar B \rightarrow
f) \over d\Gamma (B \rightarrow f) + d\Gamma (\bar B \rightarrow f)}  \\
&& \nonumber \\
& = & - Im (u_f)\;\sin (x\Gamma t).
\end {eqnarray}
Note that if we had not assumed eq.(21) we would have obtained an extra
term $(1 - |u_f|^2) \cos(x\Gamma t)$ , which could be isolated by the
time-dependence characteristic of the cosine factor.

It is however useful to consider the ``time-integrated asymmetry'' which
is formed by the rates that are integrated over all positive time, from 0
to $\infty$,
\be
a_f = - \left( {x \over 1+x^2}\right)\;\;Im\;(u_f).
\ee
This is a neat formula because it depends only on the
experimentally determinable quantities: the mixing parameter $x$ and the
CKM-matrix phase that determines $Im\;(u_f)$.

The present value of $x$ (averaging over the data of ARGUS and CLEO groups)
[3] is
\be
x =  0.67 \pm 0.10.
\ee

A comment in regard to the sign of $x$ is relevant here. In Eq.(12) we
defined $x = (m_2 - m_1)/\Gamma$, but the mass-eigenstates $B_1$ and $B_2$
were not identified. If the parameter $y$ were not zero one could,
e.g.,~identify $B_2$ as the long-lived state. But in the limit of $y = 0$,
we have to obviously look for a criterion other than the lifetime
difference: One may, for instance, define the state $B_2$ to be that which
decays dominantly into CP-odd states, assuming the CP violations to be
``small''. To be specific, we shall define the state $B_2$ to be that
which decays to the CP-odd state $J/\psi K_S$. Methods suggested [9] to
fix the sign of $x$ need the timing of kaon decay in the decays $B_d
\rightarrow J/\psi K_S$ at a $B$-Factory.  Here the idea is to exploit the
strangeness oscillations to serve as `analysers' of bottom oscillations to
decide the sign of $x$.

As for the factor $Im\;(u_f)$ in Eq.(25) we turn to the standard model with 3
generations of quarks. Unitarity of the $3 \times 3$ CKM-matrix implies the
condition (column 1 with column $3^*$)
\[
V_{ud} V^*_{ub} + V_{cd} V^*_{cb} + V_{td} V^*_{tb} = 0.
\]
This triangle condition can be recast in simpler notation as follows
\begin{eqnarray}
&W_u + W_c + W_t = 0, \\
&W_q  \equiv V_{qd} V^*_{qb},\;\;(q=u,c,t). \nonumber
\end{eqnarray}

\begin{figure}[htb]
\vspace{5truecm}
\caption{CKM Unitarity Triangle }
\end{figure}

The three angles of the triangle OAB are given by
\be
\alpha = {\rm arg}\;\left[-W_t/W_u\right],\;\; \beta = {\rm arg}\;[-W_c/W_t],
\;\;\gamma  = {\rm arg}\; [-W_u/W_c].
\ee
The area of the triangle is a measure of the CP violation in the
standard-model framework with 6 quarks.

Needless to say, it is important to establish this triangle by all
possible experiments. One has to make sure that it is not the degenerate
case of a straight line, and that the lengths of the sides are such that the
triangle does close. Customarily one chooses to orient the unitarity
triangle so that $W_c$ lies along the horizontal axis in the Argand plane.
This amounts to choosing a phase convention.  In the convention adopted by
the Wolfenstein parameterization of the CKM matrix, $W_c \simeq -
A\lambda^3$ is real and the side OB of the triangle is along the $\rho$
axis.

To keep track of the several phases that arise in writing $u_f$, we adopt the
approach of Nir and Silverman [10].  We write
\be
u =\xi\;{X \over X^*}\cdot{Y \over Y^*}\cdot{Z \over Z^*}\;,
\ee
where the first factor $\xi$ is the CP-eigenvalue $(\pm 1)$ of the state $f$.
Each of the factors $X,\;Y,\;Z$ turns out to be a product of two CKM-matrix
elements: $X$ comes from the two $V$'s at the two ends of the $W$ in the
quark-level diagram of the decay $B \rightarrow f$; for instance,
\begin{eqnarray}
X [\bar b \rightarrow \bar cc \bar s] & = & V_{cb} V^*_{cs} \\
X [\bar b \rightarrow \bar uu\bar d] & = &  V_{ub} V^*_{ud} \nonumber \\
& =& W^*_u.
\end {eqnarray}
The factor $Y$ arises from the dominant box-diagram (with virtual $t\bar t$
pair) describing $B\bar B$ mixing; it does not depend on state $f$;
it denotes the
possibility that by virtue of mixing, the decay of $B(d\bar b$) arises from an
initial $\bar B(b\bar d)$ state:
\begin{eqnarray}
Y [\bar B(b\bar d) \rightarrow (t\bar t\;{\rm box}) \rightarrow
B(d\bar b)] & = & V^*_{tb} V_{td} \nonumber \\
& = & W_t.
\end{eqnarray}
The factor $Z$ is similar to $Y$; it arises from $K\bar K$ mixing when the
final state $f$ contains a $K_S$ or a $K_L$ meson. Obviously we ought to
set $Z = 1$ if there is no neutral kaon in the state $f$. For evaluating
$Z$ we read off the $K\bar K$-mixing phase from the box-diagram involving
only $c$-quarks (note that the `box-mechanism' is relied upon only to the
extent of getting the phase). If we start with a neutral $K$ with positive
strangeness in the primary decay of $B$, we then have
\be
Z [K(\bar sd) \rightarrow (\bar cc\;{\rm box}) \rightarrow \bar K(\bar ds)] =
V_{cs} V^*_{cd}\,.
\ee

The neutral $B$ decay mode that enjoys maximum attention and is regarded
as the ``show-case sample'' for the decays in $B$ physics, is the two-body
mode
\be
B  \rightarrow J/\psi  + K_S\,.
\ee
It has the distinctive lepton-pair of the $J/\psi$ for its triggering. The
final hadronic state has $CP = -1$ because the $K$ has to be emitted in a
relative $p$-wave. The decay is governed by the CKM-favoured transition
$\bar b \rightarrow \bar cc\bar s$.  Substituting for $X,\;Y$,~and $Z$ in
Eq.(29) the argument of $u$ is given by
\begin{eqnarray}
{\rm arg}\;(u) & = & -2\;{\rm arg} (XYZ) \nonumber \\
& = & -2\; {\rm arg} [W^*_c W_t]  \nonumber  \\
& = & 2\beta, \nonumber
\end{eqnarray}
and hence the imaginary part of $u$ is
\be
Im\;[u(B_d \rightarrow J/\psi K_S)] =  + \sin (2\beta).
\ee
A measurement of the time-integrated asymmetry in this channel can
therefore determine the angle $\beta$ by means of eq.(25). The asymmetry in
respect of the decay $B_d \rightarrow D^+ D^-$
also gives information on the same angle.

Next, consider the CKM-suppressed transition  $\bar b \rightarrow \bar
uu\bar d$ which underlies the $B$ decay into the CP-even state of two pions,
\be
B_d \rightarrow \pi^+ \pi^-.
\ee
As there is no kaon in the final state we immediately set $Z = 1$, and using
eqs.(31) and (32) obtain
\begin{eqnarray}
{\rm arg}\;(u) & = & +2\;{\rm arg}\;[W^*_u W_t] \nonumber \\
& = & -2\; {\rm arg}\;[W_u / W_t] \nonumber \\
& = & -2\;(\pi - \alpha); \nonumber
\end {eqnarray}
we thus have
\be
Im\;[u(B_d \rightarrow \pi^+ \pi^-)] = + \sin(2\alpha).
\ee

It should however be pointed out that this determination of $\alpha$ is
obscured by what is usually called the ``rescattering pollution''.  This
complication arises because the $2\pi$ final state can also be reached from
a differnt decay mode by final state rescattering; for instance,
\begin{eqnarray}
&(1)& B(\bar b d) \rightarrow \bar u (u \bar d)\; d \rightarrow \pi^+ \pi^-,
\nonumber \\
&(2)& B(\bar b d) \rightarrow \bar c (c \bar d)\; d \rightarrow D^+ D^-
\rightarrow  \pi^+ \pi^-. \nonumber
\end{eqnarray}
Here the last step $D^+D^- \rightarrow \pi^+\pi^-$ proceeds by strong
interactions (say,
by the exchange of a $D^*$ meson).  The decay amplitudes can be written as
\begin{eqnarray}
A_{2 \pi}& = & W_u |A_1| e^{i \delta_1} + W_c |A_2| e^{i \delta_2},\nonumber \\
\bar A_{2 \pi} & = & W_u^* |A_1| e^{i \delta_1} + W_c^* |A_2| e^{i
\delta_2}, \nonumber
\end {eqnarray}
where the $\delta$'s are strong interaction phases. We obtain eq.(37) if
we can show that $|A_2| \ll |A_1|$. Unfortunately a reliable estimate of
$A_2$ does not exist because one does not know how to calculate the
``long-range'' contributions to the reaction $B \rightarrow \bar D D$.
Perturbative QCD calculations of Penguin diagrams are not meaningful here as
the $Q$-value involved in the decay is not large, $(m_B - 2 m_D) \simeq
1.5$~GeV.

The haze due to rescattering pollution is however minimal or absent in
determining the angle $\beta$ from the decay (34). This is because the
important ``second'' mechanism involving rescattering
\[
B \rightarrow D_s^+ D^- \rightarrow J/\psi K_S,
\]
also has the \underbar {same} weak phase as the first mechanism.

The ``third angle'' $\gamma$ will perhaps be the most difficult to
determine as one needs to measure the asymmetry in the $B_s$ decays.  The
decay that is usually suggested is $B_s \rightarrow \rho + K_S$ for which
\be
Im\;[u(B_s \rightarrow \rho^0 K_S) = + \sin (2\gamma).
\ee
Asymmetries in $B_s$ decays in general are expected to be quite small as
the arguments based on the box-diagram imply a very large $B_s-\bar B_s$
mixing (say, $|x_s| \gsim 10)$; i.e., the $B_s-\bar B_s$ oscillations
would be so rapid that whatever may be the initial state it quickly ends
up as a near equal mixture of $B_s$ and $\bar B_s$, and thus any asymmetry
in the decays gets washed out [4].  Further this means one needs a
$B_s$-Factory which is a $e^+e^-$ machine tuned for the $\Upsilon(5S)$ at
10.865~GeV; this resonance has a small production cross section $(\simeq
0.16~nb)$ and a small branching fraction $(\lsim 0.1)$ for decay into
$B_s\bar B_s$ state.

The present phenomenological analyses (based on the available data on
$B_d\bar B_d$ mixing, $|V_{ub}/V_{cb}|$, etc.) cannot
ascertain even whether the
angle $AOB$ is acute or obtuse, that is, whether the apex $A$ lies in the 1st
or the 2nd quadrant in the $\rho\eta$-plane.  For a discussion of the $B$
decay asymmetries outside the standard model framework, see Ref. 11.

{\bf Winstein Ambiguity}: This interesting ambiguity [12] stems from the
possibility that the two measured asymmetries $a(J/\psi K_S)$ and
$a(\pi^+ \pi^-$) are
equal and opposite such that the following relation is satisfied
\be
\sin (2\beta) =  - \sin (2\alpha).
\ee

This relation in fact is the prediction of the Super Weak (SW) model,
which was originally proposed by Wolfenstein in 1964 to account for the
data on CP violation in neutral kaons. It should be noted that the model
is yet to be falsified by experiments: Based on the analyses of their full
data the experimental groups have annonced at the 1991 Lepton-Photon
Symposium, Geneva,
\begin{eqnarray}
Re \left(\epsilon' \over \epsilon \right) & = & (2.3 \pm 0.7) \times 10^{-3},\;
(NA31) \nonumber \\
& = & (0.60 \pm 0.69) \times 10^{-3},\; (E731).\nonumber
\end{eqnarray}
This ratio ought to vanish in SW.  According to the SW scenario all the CP
violation effects arise only from the transitions involving $\Delta\;({\rm
flavour}) = 2$; in regard to the off-diagonal element in the $K\bar K$
mass-matrix the SW transitions make the parameter $M_{12}$ complex but
keep the $\Gamma_{12}$ real. A consequence of the SW model is that the
asymmetries for CP-odd and CP-even decay states should be equal and
opposite as implied by eq.(39).

\begin{figure}[htb]
\vspace{8truecm}
\caption{Winstein's Ambiguity}
\end{figure}

Since the measurements of asymmetries will be having some errors, one
could in practice test only the consistency of the relation (39). Hence it
is appropriate to display the implications between the CKM parameters, not
as a curve but, as a region in the parameter space. Winstein's analysis shows
(see Fig.~2) that for realistic errors on the measurements there exists a
sizable region in the $\rho\eta$-plane wherein we will
not be able to distinguish
between the SM and SW; for further details see Ref. 13. Note however that
an observation of CP violation in the $B^\pm$ sector would at once rule out
the SW option, as there is no mixing in the case of charged mesons.

\bigskip

\centerline {3.~CP ASYMMETRY IN LIKE-SIGN DILEPTON EVENTS}

\medskip

\nobreak
I am talking about this because as of now this is the only asymmetry
relating to CP violation in the B system on which an experimental number
is being mentioned. The extraction of this number of course does not need
any timing information, or reconstruction of a particular exclusive mode.
It is a measurement, at a $\Upsilon(4S)$ machine, of the difference in the
total numbers of like-sign dilepton events of both kinds, $N(++)$ and
$N(--)$.

Following the $\Delta B = \Delta Q$ rule of the SM, we define the two
allowed amplitudes for decay into an exclusive semileptonic mode $(\ell^+
X)$ and its CPT-conjugate  $(\ell^- \bar X)$ as
\begin{eqnarray}
A(\ell^+)& = &\langle \ell^+ X |T| B \rangle, \nonumber \\
A(\ell^-)& = &\langle \ell^- \bar X |T| \bar B \rangle;
\end{eqnarray}
these are related by CPT invariance:
\[
A(\ell^-) = [A(\ell^+)]^* e^{2i\delta},
\]
where $\delta$ is the eigenchannel scattering phase due to electro-weak
interactions in the final state.  Due to mixing, the time-evolved states
$B(t)$ and $\bar B(t)$ do lead to `wrong-sign' leptons.  If all the
physically connected final channels are summed over, one gets effectively,
\[
|A(\ell^+)| = |A(\ell^-)|.
\]
We now sum over the final spins and momenta, and finally sum over the
channel index $X$ to pass to the inclusive limit. In this way, without
making the assumptions of eqs. (19) and (20), we obtain the same overall
constant (which we suppress) in the following four time-dependent rates,
written for brevity as a matrix product:
\be
{d\Gamma \over dt} \left[
     \begin{array}{c} B \rightarrow \ell^+ \\
                \bar B \rightarrow \ell^- \\
                                B \rightarrow \ell^- \\
                                \bar B \rightarrow \ell^+
         \end {array} \right]
         \sim e^{-\Gamma t}\;
         \left[ \begin{array}{ll}
                 1 & 1 \\
                         1 & 1 \\
                         |g|^2 & -|g|^{2} \\
                         |g|^{-2} &  -|g|^{-2}
         \end{array} \right] \;
    \left[ \begin{array}{c}
          \cosh (y\Gamma t) \\
          \cos (x\Gamma t) \end {array} \right]\; .
\ee

Consider the semileptonic $B$ decays starting with $\Upsilon (4S) \rightarrow
B\bar B$. Taking the
electric charge of the lepton that tags the $B$ to be the first index (say),
we see that the last two equations above lead to the class of events with
likesign dileptons. From these rates the dilepton charge asymmetry [14] is
(independent of time)
\be
a(\ell\ell) = {N(++) - N(--) \over N(++) + N(--)} = {1 - |g|^4 \over 1
+ |g|^4}.
\ee
A preliminary value quoted by the CLEO group [1], on the basis of about
200 events containing likesign dileptons, is
\be
a(\ell\ell) = (- 0.58 \pm 9.9 \pm 3.8)\;\%\;.
\ee
This means, at 90\% C.L., the dilepton charge asymmetry (originating from
the CP violation in $B\bar B$ mixing) is less than 14\%;
an alternative way of stating this is
\be
Re(\epsilon) < 4 \% ,
\ee
where $\epsilon$ is given by $g = (q / p) = (1 - \epsilon) / (1 +\epsilon)$.
Note that in the case of $K_L$ decays the celebrated SLAC leptonic
charge-asymmetry implied $Re(\epsilon_K) \simeq 0.16\%$.

The LEP groups are expected [3] to provide soon a number for the asymmetry
$a(\ell\ell)$, which is a factor 2 better than the above.  However its
measurement at the level of $10^{-3}$ (which is the estimate from the models
[15]), requires about $10^{10}$ upsilons and several years of
running of a $B$ Factory (note that the number goes inversely
as the square of the asymmetry).

\bigskip

\centerline {4. ASYMMETRY TO OBTAIN $y$}

\medskip

An asymmetry which would eventually shed light on the small parameter $y$ in
eq.(12), can be constructed [9] from the time dependence of the $B_d$ and
$\bar B_d$
decays into the same CP eigenstate $f$. If $t$ denotes the time of decay
$B/\bar B \rightarrow f$ and $t'$ the lepton-tag time, then the relative
time $\tau = (t - t')$ can
be either positive or negative, whereas the sum $T = (t+t')$ must be greater
than $|\tau|$.  Hence the probability that one of the $B$'s decays into
$\ell^+$ or $\ell^-$
and the companion $B$ decays to the state $f$ after a time $\tau$ can be worked
out exactly:
\begin{eqnarray}
W_\ell^{(f)}(\tau) & = & \int^\infty_{|\tau|} d(t+t')\; \{D[\ell^+(t'),\;
f(t)]  + D[\ell^-(t'),\;f(t)]\} \\
& \sim & e^{-\Gamma|\tau|} \{(1 + |u_f|^2)\;
\cosh(y\Gamma\tau) + 2 Re(u_f)\;
\sinh(y\Gamma\tau) \nonumber \\
& & \qquad -  \Omega\; (1 - |u_f|^2)\;
\cos(x\Gamma\tau) + 2
\Omega \;
Im(u_f) \sin(x\Gamma\tau)\}, \nonumber
\end{eqnarray}
where the `overlap' parameter $\Omega$ is defined as
\be
\Omega = {1 - |g|^2 \over 1 + |g|^2}.
\ee

The corresponding time-asymmetry we construct is
\begin{eqnarray}
A_f (\tau) & = & {W_\ell^{(f)} (\tau) - W_\ell^{(f)} (- \tau) \over
W_\ell^{(f)} (\tau) + W_\ell^{(f)} (-\tau)} \\
& = & {2 Re(u_f) \sinh (y\Gamma\tau) + 2 \Omega \;Im (u_f) \sin (x\Gamma
\tau) \over (1+|u_f|^2) \cosh (y\Gamma\tau) - \Omega (1 - |u_f|^2) \cos
(x\Gamma\tau)}.
\end {eqnarray}
Here the effects due to CP violations enter through the nonvanishing of
the parameters $\Omega$, $Im(u_f)$ and $(1 - |u_f|^2)$.
In order to go to the limit of
CP conservation therefore we set these parameters zero, and use the
identity
\[
{2 Re\;u  \over 1 + |u|^2} = \;\pm 1 \;\; \mp {|1 \mp u|^2  \over 1 + |u|^2}.
\]
Since we must have $u_f = \pm 1$ according as the state $f$ is CP-even or
CP-odd, we obtain for the asymmetry
\be
A_f(\tau) = \pm \tanh (y\Gamma\tau),\;\;{\rm where\,CP}\;(f) = \pm (f).
\ee

This is a useful formula. The corrections to it due to violations of
CP-symmetry arise from terms which are of {\it second- and higher-order
smallness}, and hence ignorable. For small enough $(y\Gamma\tau)$ and,
say, for an odd CP state such as $J/\psi  K_S$, we see
\be
A_f(\tau) = - \tanh (y\Gamma\tau) \simeq  (\Gamma_1 - \Gamma_2)\; \tau;
\ee
hence the sign of $y$ gets determined if we regard the state $B_2$ to be
dominantly CP odd.

Note that a measurement of the asymmetry $A_f(\tau)$ at a $B$ Factory needs no
special experimental effort; one needs to reanalyse the same data which
will be collected for studying CP violation in the decays of $B$ into CP
eigenstates, such as in (34) and (36). This is because for the
CP-asymmetries one looks at the difference in the rates of $\ell^+$ and
$\ell^-$ emission in correlation with a CP-eigenstate $f$, while for
$A_f(\tau)$ one looks at the corresponding sum as in eq.(45). Secondly, as
the formula (50) is the same for all decay states having the same
CP-eigenvalue, we can easily enlarge the event sample.

It should be mentioned that if we start by defining eq. (47) in terms of
the time-integrated rates $\int^\infty_0 d\tau W(\pm \tau)$, the corresponding
time-integrated asymmetry will be given by
\[
A_f = (\pm)\;y,
\]
where the sign $(\pm)$ refers to the CP signature of the state $f$.

\bigskip

\centerline {5.~TEST OF THE $\Delta B = \Delta Q$ RULE}

\medskip

Recently Kobayashi and Sanda [16] have suggested ways for testing CPT
invariance at a $B$-Factory. Their program to examine the mass-matrix CPT
violation $[(q_1/p_1) \neq {(q_2/p_2)}]$ supposes that both CPT
invariance and the
$\Delta B = \Delta Q$ rule are respected by the amplitudes of the
semileptonic $B$ decays.

We shall take a different stand [17]: We take CPT invariance as sacrosanct
and explore the breakdown of the Standard Model through violations of the
$\Delta B = \Delta Q$  rule in neutral $B$ decays.
According to this selection rule the
transition to an exclusive semileptonic state will be classified as
`allowed' and `suppressed' as follows:

\medskip

\[
\begin{tabular}{|l|l|}
\hline
Allowed & Suppressed \\
\hline
&\\
$B \rightarrow R \ell^+:\,A$ & $\bar B \rightarrow R \ell^+:\, \rho A$ \\
$\bar B \rightarrow \bar R \ell^-:\,\bar A$ & $B \rightarrow \bar R \ell^-:
\,\bar\rho\bar A $\\
&\\
\hline
\end{tabular}
\]

\medskip

\noindent here the symbol $R$, stands for the remaining part of the final state
including the appropriate neutrino, e.g., $R$ could stand for $D^{*-}
\nu_\ell$. The two complex parameters $\rho$ and $\bar \rho$ (both of
which ought to  vanish in SM) are related by CPT invariance
\be
\bar\rho  = \rho^*.
\ee

Let us consider (at an asymmetric $B$ Factory) the "timed" unlike-sign
dilepton events from exclusive decays of the $B_d$. We denote their number as
$N[R \ell^+(t'),\; \bar R \ell^-(t)]$. Defining the variables $T = (t + t')$
and $\tau = (t - t')$, we integrate over $T$ from $|\tau|$ upto $\infty$,
and obtain (suppressing an irrelevant overall factor)
\begin{eqnarray}
&N[R\ell^+,\,\bar R\ell^-;\; (\tau)] \sim e^{-\Gamma |\tau|} \; [|\gamma|^2
e^{y\Gamma\tau} + e^{-y\Gamma\tau} + 2\,Re (\gamma e^{ix\Gamma\tau)}]  \\
&N[R\ell^+,\,\bar R\ell^-;\;(-\tau)] \sim e^{-\Gamma |\tau|} [|\gamma|^2
e^{-y\Gamma\tau} + e^{y\Gamma\tau} + 2\,Re (\gamma e^{-ix\Gamma\tau)}]
\end {eqnarray}
the two rates differ by the sign of $\tau$; the parameter $\gamma$  is
\be
\gamma = {1 + (\bar \rho /g) \over 1 - (\bar \rho /g)}\;\; {1 - \rho g \over
1 + \rho g}.
\ee

The asymmetry that should vanish according to the $\Delta B = \Delta Q$
rule can therefore be constructed
\be
A_{BQ}(\tau) = {N[R\ell^+, \bar R\ell^-;\;(\tau)] - N[R\ell^+,\,\bar
R\ell^-( -\tau)] \over
N[R\ell^+, \bar R\ell^-;\;(\tau)] + N[R\ell^+,\,\bar
R\ell^-( -\tau)]}.
\ee
Integrating the $N$'s over $\tau$ from 0 to $\infty$, the corresponding
time-integrated asymmetry becomes
\be
a_{BQ} = {1 \over 2 + x^2 - y^2} \;\left[-Im (\gamma) x (1 - y^2) - {1
\over 2}  (1 - |\gamma|^2) \; (1 + x^2)\,y\right].
\ee

Assuming CPT invariance, we substitute eq.(51) in eq.(54) and observe that
the only phase which controls the phase of $\gamma$ is that of $(g\rho)$
which is a
phase-convention-independent combination. For small values of rho, since
\be
\gamma \simeq  1 - 4i\, Im(g\rho),
\ee
we see that, by ignoring terms which are second and higher order in $\rho$,
eq.(57) becomes
\be
a_{BQ} \simeq \left[{4x (1 - y^2) \over 2+x^2 - y^2}\right]\;Im(g\rho).
\ee
The first factor is harmless because its value is about 1.1 for $x \simeq 0.67$
and $y \simeq 0$. Thus this asymmetry directly measures
the quantity $Im(g \rho)$. This essentially is $Im(\rho)$ if
the CP violation at the "structure-level" turns out to be negligibly
small, $g \simeq 1$, as is indicated by several model estimates [15].
In any case as the selection rule $\Delta B = \Delta Q$ is an integral
part of the
standard model and plays a crucial role in $B$ Physics, an experimental
test of its validity is obviously important.

To conclude, after the pion Factories LAMPF and TRIUMF, the
recently-funded Frascati Phi-Factory (DA$\phi$NE) and the Spain's~Tau-Charm
Factory --- the next step seems to be a $B$-Factory. The 1987 discovery of the
ARGUS group that there is sizable mixing in $B_d$ mesons, generated great
interest in $B$-physics. The $B$'s may indeed be the heaviest quark mesons
which permit an experimental study of the origin of CP violation (the top
bound states would decay extremely rapidly by the kinematically allowed
transition $t \rightarrow b W^+$). Against such a background it is difficult to
explain why, as of now, none of the proposals (CESR-B, KEK-B, PEP-II,
CERN-ISR, $\ldots$) has been funded.  However in view of the tremendous
enthusiasm and the world-wide support that this project seems to enjoy, it
will be fair to predict that a $B$-Factory will be built --- the only question
is, where and when.

I would like to acknowledge several helpful discussions I had with \break
G.~V.~Dass during the preparation of the talk.  I also thank him for a
careful reading of the manuscript.

\newpage

\begin{center}\bf REFERENCES \\
\end{center}

\begin{enumerate}

\item P. Drell, Intern. Conf. of HEP 1992, Dallas.

\item K. Lingel et al, Cornell preprint CLNS 91-1043 (1991).

\item  V. Sharma, X DAE HEP Symposium 1992, Bombay.

\item  J. F. Donoghue, E. Golowich and B.R. Holstein, DYNAMICS OF THE
          STANDARD MODEL, (Cambridge,1992).

\item  A. B. Carter and A. I. Sanda, Phys. Rev. Lett. 45, 952 (1980);
          Phys. Rev. D23,1567 (1981).

\item I. I. Bigi and A. I. Sanda, Nucl. Phys.B193, 85(1981)

\item I. Dunietz and J. L.Rosner, Phys. Rev. D34,1404 (1986).

\item  I. I. Bigi, V. A. Khoze, N. G. Uraltsev and A. I. Sanda,
          in CP VIOLATION, ed. C. Jarlskog (World Scientific,1989), p.175;
I. Dunietz, in B DECAYS, ed. S. Stone (World Scientific, 1992), p.393.

\item  G. V. Dass and K. V. L. Sarma, Int. J. Mod. Phy. 7, 6081 (1992);
(E) \underbar {8},~1183 (1993).

\item  Y. Nir and D. Silverman, Nucl. Phy. B345, 301 (1990).

\item  C. O. Dib, D. London and Y. Nir, Int. J. Mod. Phys. 6, 1253
          (1991); Y. Nir and H.R. Quinn, Ann. Rev. Nucl. Part. Sci.
\underbar{42}, 211 (1992).

\item  B. Winstein, Phys. Rev. Lett. 68, 1271 (1992).

\item  J. M. Soares and L. Wolfenstein, Phys. Rev. D47, 1021 (1993).

\item  L. B. Okun, V. I. Zakharov and B. M. Pontecorvo, Lett. Nuo. Cim.
          13, 218 (1975).

\item  J. Hagelin, Nucl. Phys. B193, 123 (1981); T. Altomari,
          L. Wolfenstein and J. D. Bjorken, Phys. Rev. D37, 1860 (1988);
          L. M. Sehgal and M. Wanninger, Phys. Rev. D42, 2324 (1990).

\item M. Kobayashi and A. I. Sanda, Phys. Rev. Lett. 69, 3139 (1992).

\item G. V. Dass and K. V. L. Sarma, to be published.
\end{enumerate}

\end{document}